\documentstyle[aps]{revtex}
\begin{document}
\def\be{\begin{equation}}
\def\ee{\end{equation}}
\def\bea{\begin{eqnarray}}
\def\eea{\end{eqnarray}}
\renewcommand{\thefootnote}{\fnsymbol{footnote}}

\twocolumn[\hsize\textwidth\columnwidth\hsize\csname@twocolumnfalse%
\endcsname

\title{Two loop results from the derivative expansion of the blocked action}
\author{Alfio Bonanno$^{(1,3)}$ and Dario Zappal\`a$^{(2,3)}$}
\address{
 $^{(1)}$Istituto di Astronomia, Universit\`a di Catania\\
Viale Andrea Doria 6, 95125 Catania, Italy\\
 $^{(2)}$Dipartimento di Fisica, Universit\`a di Catania, 
$^{(3)}$INFN, sezione di Catania\\
Corso Italia 57, 95129 Catania, Italy\\
}

\date{\today}
\maketitle
\draft
\begin{abstract}
A derivative expansion of the Wegner-Houghton equation is derived 
for a scalar theory. The corresponding full non-perturbative
renormalization group equations 
for the potential and the wave-function renormalization function
are presented. We also show that the two loop perturbative anomalous
dimension for the $O(N)$ theory can be obtained by means of a polynomial 
truncation in the field dependence in our equations.
\end{abstract} 

\pacs{pacs 11.10.Hi , 11.10.Gh \hfill PREPRINT INFNCT/6-97}
]

In field theory the description of physical observables 
at energy scales that can be several orders of 
magnitude below the cut-off scale, is one of the most relevant issues.
This problem is studied by means  of the renormalization group (RG)
and, in particular, the Kadanoff-Wilson formulation of the 
RG \cite{wilson}, based on the blocking procedure,
is probably the most preferable starting point for any application.
The Wegner-Houghton equation \cite{wegner} is a specific realization 
of the blocking procedure that relates the action defined at the scale 
$k$ to the action at the scale $k-\delta k$ through a functional integration 
over the infinitesimal 
momentum shell $\delta k$, for $\delta k \to 0$. This differential equation
is exact, as shown in \cite{wegner} since it contains the whole 
effect due to the integration of the fast modes. However it has no known 
analytical solution and one has to resort to approximations in order to deal 
with a more tractable problem.

The most promising approach consists in a derivative expansion
where the action is expressed as a sum of terms containing an increasing 
number of field derivatives. The lowest order corresponds to a momentum 
independent projection of the Wegner-Hougton equation which is then reduced
to a renormalization group equation for the local potential \cite{hasen}. 
To the next order, $O(\partial^2)$ terms are included and one gets two coupled
differential equations for the potential $V$ and the wave-function 
renormalization function $Z$. In \cite{morris} and, more recently, 
in \cite{morris2},
these equations are studied
numerically to investigate fixed points and determine critical exponents.
Some applications of the $O(\partial^2)$ truncation are considered 
in \cite{polonio,greiner,noi,noi2}.

Here we focus on the connection between 
perturbative calculations and the various approximations of the Wegner-Houghton
equation. The exactness of the full equation  implies that one should be able 
to recover the  perturbative results order by order in the coupling constant 
at least in some approximation scheme. For instance if  one considers a 
polynomial expansion in the field $\phi$ for the potential $V$ of a scalar 
theory and reduces the equation for $V$ to a set of differential equations 
for the various polynomial coefficients ($m^2,\lambda, ...$),
it is very easy to check that the equation for the quartic coupling $\lambda$
yields the perturbative one loop $\beta$-function,
provided that one neglects all couplings corresponding to non-renormalizable 
powers of $\phi$ ($\phi^6,\phi^8,...$) 
and takes the ultraviolet (UV) limit where
the mass is negligible compared to the running scale \cite{pollo}.
However, going beyond one loop is no longer straightforward.

In this paper we specifically consider the anomalous dimension $\eta$
for a $O(N)$ symmetric scalar theory, which gets the first non-vanishing 
contribution at two loop, and show how the $O(\lambda^2)$ term is obtainable 
from the Wegner-Houghton equation.
In fact the two loop anomalous dimension and $\beta$-function have been 
derived in \cite{wett1,wett2,wett3}, by making use of the average action 
formalism; 
however this result is not entirely satisfactory since the authors, by 
employing that formalism, are forced to introduce a smooth cut-off, namely an 
exponentially decreasing function, in order to integrate out the UV modes. 
Instead, in the following it is shown that the very simple and intuitive 
scheme which yields the one loop $\beta$-function can be extended to deduce 
$\eta$ at two loop level, just solving the differential equation obtained 
for a particular truncation of the polynomial expansion of $Z$, not resorting 
to any smooth cut-off.

Moreover, starting from the two loop anomalous dimension, we compare 
expansions of the action having different nature and leading to different 
equations for $Z$. We show that the plain derivative expansion introduced in 
\cite{novik,zuk} has the correct UV behavior, whereas the one 
suggested by Aitchison and Fraser \cite{fraser} is misleading in the UV region.
As a consequence we can discriminate between the two approaches and choose the
first one in order to consider any further application.

Let us consider the Euclidean $N$-component scalar theory ($a,b=1,2,..N$ and
$\phi^2=\phi_i \phi^i$) 
in four dimensions, and, according to our choice of considering 
a derivative expansion of the action, truncated at $O(\partial^2)$, the 
action at the scale $\Lambda$ reads 
\bea\label{bareact}
&&{S}_{\Lambda}[\phi]=
\int d^4x \Big (\frac{1}{2}
{Z}^{ab}\partial_\mu\phi_a\partial^\mu\phi_b+V\Big )
\eea
where the potential $V$ and the wave-function renormalization 
function $Z_{ab}$ are 
polynomial in $\phi_i$. In the following the $O(N)$ symmetry of the theory
is not required, except when explicitly indicated.
The blocked action $S_{k}[\Phi]$, which is the action at the scale 
${ k}<\Lambda$ and which is a functional of the field $\Phi(x)$ 
that, by construction has Fourier components only up to the scale 
$k$, is obtained by functional integration from 
$S_{\Lambda}[\phi]$. Up to terms that are not relevant for the Wegner-Houghton 
equation, the functional integration yields 
\bea\label{blockact}
&&{S}_{k}[\Phi]=
{S}_{\Lambda}[\Phi]+\frac{1}{2}
{\rm Tr}'{\rm ln} \left( K_{ij} -\delta_{ij} K_0 \right )
\eea
where 
$K_{ij}=(\delta^2 S_{\Lambda}/\delta\phi_i\delta\phi_j)|_{\phi=\Phi}$
and ${\rm Tr}'$ is the trace of the internal degrees of freedom and of the 
modes within the shell $ k-\Lambda$, and 
finally $\delta_{ij}K_0$ is the free propagator with mass $m_0$.

The Wegner-Houghton equation is the 
differential equation for ${S}_{k}[\Phi]$ obtained by 
differentiating 
Eq.(\ref{blockact}) with respect to $k$, with 
the boundary condition fixed by ${S}_{\Lambda}[\phi]$.
By double-differentiating Eq.(\ref{bareact}) we get
\bea\label{cappaij}
&&{K}_{ij}(x,y)=
\nonumber\\
&&\left [\frac{1}{2}{Z}_{ab{\textstyle ,}ij}(x)
\frac{\partial \Phi^a(x)}{\partial x_\mu} 
\frac{\partial  \Phi^b(x)}{\partial x^\mu}
+V_{{\textstyle ,}ij}(x)\right]\delta^4(x-y)
\nonumber\\
&&+\int d^4 x'\Biggl [\delta^4 (x-x') Z_{(bj){\textstyle ,}i}(x')
\frac{\partial\Phi^b(x')}{\partial{x'}^{\mu}}
\frac{\partial}{\partial{x'}_{\mu}}\delta^4(x'-y)
\nonumber\\
&&+\delta^4 (x'-y) Z_{(ib){\textstyle ,}~j}(x')
\frac{\partial \Phi^b(x')}{\partial{x'}^{\mu}}
\frac{\partial}{\partial{x'}_{\mu}}\delta^4(x-x')
\nonumber\\
&&+Z_{(ij)}(x')
\frac{\partial}{\partial{x'}^{\mu}}\delta^4 (x-x')
\frac{\partial}{\partial{x'}_{\mu}}\delta^4 (x'-y)\Biggr]
\eea
where the notations $A_{(ij)}=(A_{ij}+A_{ji})/2$ and $A_{ij{\textstyle ,}k}=
\partial A_{ij}/\partial \Phi_k$ have been used.
A comment is in order: in deriving Eq.(\ref{cappaij}) no integration by parts
has been performed and thus no surface term in the action has been neglected;
indeed these terms, which are irrelevant in the equation of motion,
supply a higher order (in the weak coupling perturbative expansion) 
contribution to Eq.(\ref{blockact}). 

By employing the conventions $\delta^4(x-y)=\langle x|y\rangle$, 
$(\partial/\partial x_{\mu})\delta^4(x-y)=-i\langle x| \widehat p^\mu
|y \rangle
=-(\partial/\partial y_{\mu})\delta^4(x-y)$
and the commutation rules $[\widehat p^\mu,f(x)]=i\partial^\mu f(x)$,
$[\widehat p^2,f(x)]=2i\partial^\mu f(x) \widehat p_\mu- \partial^2 f(x)$,
$K_{ij}(x,y)$ can be cast in the operatorial form  
$\langle x|K_{ij}|y \rangle $ with 
\bea\label{cappaijop}
&&K_{ij}=Z_{(ij)}\widehat p^2+V_{{\textstyle ,}ij}+
i{{X_i}^b}_j\partial^\mu\Phi_b \widehat p_\mu
\nonumber\\
&&+\frac{1}{2}Z_{ab{\textstyle ,}ij}\partial_\mu\Phi^a\partial^\mu\Phi^b-
\partial_\mu\left(Z_{(ib){\textstyle ,}~j}\partial^\mu\Phi^b\right )
\eea
where we have ordered all coordinate operators to the left of the momentum 
operators and 
\be
{{X_{i}}^b}_j=
{{Z_{(i}}^{b)}}_{{\textstyle ,}~j}-
{{Z_{(j}}^{b)}}_{{\textstyle ,}i}+
{Z_{(ij)}}^{{\textstyle ,}b}.
\ee
Next step concerns the evaluation of the trace in Eq.(\ref{blockact}).
An expansion of the trace in derivatives of the fields is required
in order to determine an evolution equation for $V$ and $Z_{ab}$,
and such a suitable expansion has been already derived  in \cite{novik,zuk}.  
It is obtained by defining $K_{ij}(u)=K_{ij}+u\delta_{ij}$
and $K_0(u)=K_0+u$, where $u$ is 
a fictitious ``mass", then deriving with respect to $u$ and finally 
expanding the derivative of the logarithm.
By introducing
\bea\label{gammau}
\Gamma(u)= 
\frac{1}{2} {\rm Tr}'{\rm ln} \left ( K_{ij}(u) - \delta_{ij} K_0(u) \right ),
\eea
the required term in Eq.(\ref{blockact}), $\Gamma(0)$, is given by
\bea\label{gammazero}
&&\Gamma(0)=-\int_0^\infty du \frac {d}{du}\Gamma(u)\nonumber\\ 
&&=-\frac{1}{2}\int_0^\infty du~{\rm Tr}' \left ( K^{-1}_{ij}(u) - \delta_{ij}
K_0^{-1}(u) \right )
\eea
Note that since the trace is performed on a finite momentum shell
no IR or UV divergence affects Eq. (\ref{gammazero}). Then we split 
$K_{ij}(u)=A_{ij}+B_{ij}+C_{ij}$ with 
\bea
&&A_{ij}=Z_{(ij)}p^2+V_{{\textstyle,}ij}\;\;\;\;\;
B_{ij}=Z_{(ij)}(\widehat p^2-p^2)\nonumber\\
&&C_{ij}=i{{X_i}^b}_j\partial^\mu\Phi_b \widehat p_\mu\nonumber\\
&&D_{ij}=\frac{1}{2}Z_{ab{\textstyle ,}ij}\partial_\mu\Phi^a\partial^\mu\Phi^b-
\partial_\mu\left(Z_{(ib) {\textstyle ,}~j}\partial^\mu\Phi^b\right )
\eea
where the eigenvalue $p_\mu$ of the momentum ( $\widehat p_\mu|p
\rangle=p_\mu|p \rangle$) 
has been introduced. Finally $K^{-1}_{ij}(u)$ in Eq. (\ref{gammazero})
can be expanded according to  
\bea\label{expans}
&&K^{-1}(u)=A^{-1}-A^{-1}(B+C+D)A^{-1}
\nonumber\\
&&+A^{-1}(B+C+D)A^{-1}(B+C+D)A^{-1}-...
\eea
For a detailed analysis of the validity of this operatorial 
expansion we address to the Report of R. Ball \cite{ball}.

In order to deduce the evolution equations for $V$ and $Z_{ab}$,
it is sufficient to retain respectively all terms with no  or two derivatives.
Actually $C$ and $D$ contain respectively one and two field 
derivatives; then in the trace in Eq. (\ref{gammazero}) it is convenient to 
order all operators with all $B$'s on the right side since, by definition,
$B|p \rangle=0$. 
Therefore B appears in the trace only inside commutators and, due 
to the commutation rules, each commutator containing $B$ provides one or two 
derivatives. It follows that the truncation displayed in Eq. (\ref {expans})
is sufficient for our purpose. As shown in \cite{zuk,fraser} the trace is then 
reduced to an integration over the spatial coordinates as well as over the 
momentum modes in the $k -\Lambda$ shell.
At this level it is allowed to integrate by parts and neglect the surface 
terms since derivatives of $\delta$ are no longer present; making use 
of $(\alpha=1/(16\pi^2)~)$
\be\label{reduction}
\int \frac{d^4 p}{(2\pi)^4}~p_\mu p^\nu~ g(p^2)=
\alpha\int dp^2~\frac{p^4}{4}~g(p^2)~\delta_{\mu}^\nu~~~~,
\ee
we finally get 
\bea\label{mostro}
&&S_{k }=S_{\Lambda} 
+\frac{\alpha}{2}{\rm tr} \int' dp^2~p^2\int d^4 x \Biggl\{
\Bigl [ {\rm ln}(Z_{(ij)} p^2+V_{{\textstyle ,}ij}) \nonumber\\
&&-{\rm ln}(p^2+m^2_0)\delta_{ij} \Bigr ]
-\int^{\infty}_0 du~ T^{ab}_{ij} \partial^\mu\Phi_a\partial_\mu \Phi_b\Biggr \}
\eea
where the trace only concerns the  matrix indices $i$ and $j$, 
$\int'$ indicates 
that the integral is restricted to the shell $k^2-\Lambda^2$ and 
\bea\label{most}
&&T^{ab}=-\frac{1}{2} A^{-1}Q^{ab}A^{-1} 
-A^{-1}M^a A^{-1}Z A^{-1}M^b A^{-1} 
\nonumber\\
&&+ A^{-1}M^a A^{-1} R^{b}A^{-1} 
+A^{-1} (R^a+S^a-X^a)A^{-1}M^b A^{-1}
\nonumber\\
&&+\frac{1}{2}A^{-1}S^{a}A^{-1}X^b A^{-1} p^2
+\frac{1}{2}A^{-1}X^a A^{-1}Z A^{-1} M^b A^{-1} p^2
\nonumber\\
&&-A^{-1}S^{a}A^{-1}Z A^{-1} M^b A^{-1} p^2
-\frac{1}{4}A^{-1}X^a A^{-1}X^b A^{-1} p^2
\nonumber\\
&&+A^{-1}M^a A^{-1}Z A^{-1}Z A^{-1} M^b A^{-1} p^2
\nonumber\\
&&-\frac{1}{2}A^{-1}M^a A^{-1}Z A^{-1} X^b A^{-1} p^2
\eea
with ${M_{kl}}^a=({Z_{(kl)}}^{{\textstyle ,}a}p^2+{V_{{\textstyle ,}kl}}^a)$
and ${Q^{ab}}_{kl}={Z^{ab}}_{{\textstyle ,}kl}$  
and ${{R_{k}}^a}_l={{Z_{(k}}^{a)}}_{{\textstyle ,} l}$ 
and ${S_{kl}}^a={Z_{kl}}^{{\textstyle ,} a}$.
It is understood that in the r.h.s. of Eq. (\ref{most}), 
the omitted lower indices  are contracted, according to the given sequence, 
in such a way that, with $a$ and $b$ fixed, each term of the sum,
as well as $T^{ab}$, is a matrix with lower indices $i$ and $j$.

The equation for the potential is then recovered from the piece not containing 
field derivatives whereas the $\partial_\mu \Phi_a\partial^\mu \Phi_b$
coefficient gives the equation for $Z_{ab}$.

For a general $N$-component theory it is not possible to commute the matrices
in Eq. (\ref{mostro}) and this makes the integration of the variable
$u$, which appears only in $A^{-1}$, a hard task. Beside a direct 
numerical approach there are special cases where Eq. (\ref{mostro}) can 
be drastically simplified. For instance in ref. \cite{zuk} it is shown how to 
perform  the integral through a change of variables if $Z_{ab}=\delta_{ab}$
in the r.h.s. of Eq. (\ref{mostro}); however this is equivalent to neglect 
all $Z_{ab}$ contributions to the scale evolution of $Z_{ab}$ which would be
governed by the potential only.

In the following we consider two special cases in which Eq. (\ref{mostro})
can be easily treated.
The first one consists in checking that for the $O(N)$ symmetric theory,
the perturbative two loop anomalous dimension is recovered from 
Eq. (\ref{mostro}). To this aim it is sufficient to truncate the general 
expansion of $Z_{ab}$ in powers of $\Phi$ up to $O(\Phi_i\Phi_j)$
\bea\label{zeta}
Z_{ab}=z_0\delta_{ab}+z_1 \Phi^2\delta_{ab}+z_2\Phi_a\Phi_b
\eea
and the potential to order $O(\Phi^4)$: 
$ V=m^2\Phi^2/2+\lambda\Phi^4/4!$.
In order to calculate the anomalous dimension, the function $z_0(k)$
must be determined. Therefore one has to retain in Eq. (\ref{mostro})
the numerical coefficient of 
$\delta_{ab}\partial^\mu\Phi^a\partial_\mu\Phi^b$ only,
which corresponds to set $\Phi=0$ in the coefficient of 
$\delta_{ab}\partial^\mu\Phi^a\partial_\mu\Phi^b$. 
In this case the $u$-integration is easily performed. 
According to Eqs. (\ref{mostro}), (\ref{zeta}) and by assuming,
as boundary conditions of our problem,
$Z_{ab}=\delta_{ab}$ at the scale $\Lambda$,
we get
\be\label{zetazero}
z_0=1+\alpha \int' d p^2~p^2 ~\frac {N z_1+z_2}{z_0 p^2+m^2}
\ee
In order to solve Eq. (\ref{zetazero}) the corresponding equations
for $z_1,z_2$ are needed. However as it will be clear at the end, 
instead of considering the full equations for these two couplings,
it is sufficient to 
retain the contributions proportional to the quartic coupling squared 
$\lambda^2$. This is achieved by taking $Z_{ab}=\delta_{ab}$
in the r.h.s. of Eq. (\ref{mostro}) 
and collecting terms proportional to 
$\delta_{ab}\Phi^2\partial^\mu\Phi^a\partial_\mu\Phi^b$ 
and $\Phi_a\Phi_b\partial^\mu\Phi^a\partial_\mu\Phi^b$. We get
\bea\label{zetadue}
&&z_1=\frac{2\alpha \lambda^2}{9}\int' d p^2~p^2 
\left [\frac {1}{3(p^2+m^2)^3}- \frac{p^2}{4(p^2+m^2)^4}\right ]
\nonumber\\
&&z_2=\frac{(N+6)\alpha\lambda^2}{9}
\nonumber\\
&&\times\int' d p^2~p^2\left [\frac {1}{3(p^2+m^2)^3}- 
\frac{p^2}{4(p^2+m^2)^4}\right ]
\eea
Eqs. (\ref{zetazero}), (\ref{zetadue}) can be solved perturbatively
order by order. We start with all couplings fixed by the boundary conditions at 
$\Lambda$, i.e. $z_0=1,~z_1=z_2=0$, put these
values in the r.h.s. of the equations and solve them, obtaining the first
order solution for $z_0, z_1,z_2$. Then the first order couplings can 
be put in the r.h.s. of the equations
to get the second order solution and so on. By this procedure we get 
to the first order, in the UV region ($k^2>>m^2$):
$z_0=1$, $z_1=(\alpha \lambda^2)/(54~ k^2)$, 
$z_2=(N+6)(\alpha\lambda^2)/(108~k^2) $. 
Effects $O(k^2/\Lambda^2)$ are neglected. 
To the second order for $z_0$ we get 
$z_0=1-(\alpha^2\lambda^2(N+2)/18){\rm ln}(k/\Lambda)$
and the corresponding anomalous dimension 
($t={\rm ln}(k/\Lambda)$)
\be\label{diman}
\eta=-\frac{\partial}{\partial t} 
{\rm ln}~z_0=\frac{\alpha^2\lambda^2(N+2)}{18}
+O(\lambda^3)
\ee
is in agreement with the standard result \cite{zinn}.
It is easy to realize now that, since the neglected terms in the differential 
equations for $z_1$ and $z_2$ are proportional to non-renormalizable 
couplings,
which are zero to the lowest order, they can only provide $O(\lambda^3)$ 
contributions to $z_0$ and to the anomalous dimension.

This result not only shows that the particular truncation adopted here 
still retains the features of the two loop perturbative calculations but 
mainly shows that the presence of the sharp cut-off $k$ 
in Eq. (\ref{mostro}) does not affect quantities such as the 
anomalous dimension which is related to a two loop diagram characterised 
by overlapping divergences.  
It follows that all difficulties raised in 
\cite{wett1} due to the use of a sharp cut-off and the necessity of 
introducing a smooth cut-off in the framework of the average action,
disappear when the sharp cut-off is embedded in the formalism developed above.

As a second example we consider Eq. (\ref{mostro}) for the $N=1$ theory.
In this case there are no more matrices and the $u$-integration is trivial.
Thus the Wegner-Houghton equation in this case reads (here $A=Zk^2+V''$ 
and the prime indicates derivative with respect to $\Phi$)
\bea\label{sist1}
&&k\frac{\partial V_{k}}{\partial k}=
-\alpha k^4
~{\rm ln}\left ( \frac{A}{k^2+m_0^2}\right )
\eea
\bea\label{sist2}
&&k\frac{\partial Z_{k}}{\partial k}=
-2\alpha k^4
\Bigl ( \frac{Z''}{2A}
-\frac{Z'A'}{A^2}
-\frac{{Z'}^2 k^2}{8A^2}
\nonumber\\
&& +\frac{Z{A'}^2}{3A^3}
+\frac{ZZ'A' k^2}{3A^3}
-\frac{Z^2{A'}^2 k^2}{4A^4} \Bigr )
\eea
These two non-perturbative
coupled equations are suitable for a numerical investigation of the 
fixed point structure of the theory as explained in \cite{morris,morris2}.

We shall now discuss 
the difference between Eqs.(\ref{sist1}), (\ref{sist2}), 
and the analogous equations
which are derived by resorting to the logarithm expansion proposed in 
\cite{fraser}. This is an expansion of different kind, as stressed by
 \cite{ball}, based on the presence of
``small" fluctuations of the field $\widetilde\Phi(x)$
around a constant background $\Phi_0$; this allows to expand the logarithm 
in Eq. (\ref{blockact}) up to some chosen power of  
the fluctuation field \cite{fraser,ball}.
According to this expansion the differential equations for $V$ and $Z$
have been derived in \cite{noi2}.
Although near the Gaussian fixed point the linearised form of 
Eq.(\ref{sist2}), and of Eq.(7) of ref.\cite{noi2} are the
same, some differences appear at the quadratic level.
To illustrate this point in a specific case it is sufficient to 
examine the equations for $Z$ when only the effects arising from the
potential are taken into account.
In this case one gets ( \cite{polonio,noi2})
\bea\label{zetafr}
&&k\frac{\partial Z_{k}}{\partial k}=
-\alpha k^4 (V''')^2 {V''\over (k^2+V'')^4}
\eea
while from Eq. (\ref{sist2}) one instead finds
\bea\label{zetareduced}
&&k\frac{\partial Z_{k}}{\partial k}=
-{\alpha k^4 (V''')^2\over 6}
{k^2 + 4V'' \over (k^2+V'')^4 }
\eea
One notes that the UV behavior is different in the two cases. It is
remarkable that this difference is responsible for the non-zero
anomalous dimension obtained above. 
Conversely the expansion due to
Aitchison and Fraser leads to a zero anomalous dimension to $O(\lambda^2)$
and the inclusion of higher order terms in the derivative expansion,
such as $O(\partial^4)$ or $O(\partial^6)$, does not change this result.
One can also check that the r.h.s. of the two equations behave very differently 
near the region of the mass gap, thus predicting different scaling 
behavior even at the crossover.
Only when one integrates the modes, in the independent mode approximation,
from $\Lambda$ to $k<<m$, Eqs. (\ref{zetafr}), (\ref{zetareduced})
differ by $O(k^4/m^4)$ and as already noted in 
\cite{polonio,zuk,ball,liao}, by
integrating up to $k=0$, one gets in both cases
\be
Z_{k=0}=1+\frac{(V''')^2}{192\pi^2 V''}   
\ee

On the basis of our result in Eq.(\ref{diman}),
we argue that the derivative expansion illustrated above is more suitable
for deriving a quasi-local approximation of the Wegner-Houghton including 
$O(\partial^{2n})$ terms in the blocked action. 

In conclusion we have derived two non-perturbative coupled differential 
equations for the potential $V$ and the wave-function renormalization 
function $Z$,
by making a derivative expansion of the blocked action and employing a sharp 
cut-off. These equations when compared to the usual perturbative result show a 
good UV behavior providing the correct two loop anomalous dimension
for the $O(N)$ symmetric scalar theory.

\acknowledgments
The authors have benefited from discussions with Sen-Ben Liao,
Janos Polonyi and Martin Reuter.

\end{document}